\documentclass[final,1p,times]{elsarticle}

 \usepackage{epsfig}

\usepackage{amssymb}
 \usepackage{amsthm}
 \usepackage{graphicx}
\usepackage{amsmath,amssymb}
\usepackage{amsfonts}
\usepackage{physics}
\usepackage{xspace}
\usepackage{bm}
\usepackage{mathrsfs}
\usepackage{ulem}

\usepackage[dvipsnames]{xcolor}
\usepackage[unicode]{hyperref}
\hypersetup{colorlinks=true, citecolor=MidnightBlue, linkcolor=CornflowerBlue, urlcolor=CornflowerBlue, linktocpage=true}
\usepackage[all]{hypcap}
\usepackage{multirow}
\usepackage[format=plain,justification=RaggedRight,labelsep=endash,font=small,labelfont=sc]{caption}

\usepackage[colorinlistoftodos]{todonotes}

\biboptions{sort&compress}

\journal{Physics Letters B}

\begin{document}

\begin{frontmatter}



\title{Stability constraint for spin equation of state
}


\author[inst1]{Asaad Daher}
\ead{asaad.daher@ifj.edu.pl}
\affiliation[inst1]{Institute of Nuclear Physics Polish Academy of Sciences, PL-31-342 Krakow, Poland}

\author[inst2]{Wojciech Florkowski}
\ead{wojciech.florkowski@uj.edu.pl}

\affiliation[inst2]{Institute of Theoretical Physics, Jagiellonian University, PL-30-348 Krakow, Poland}

\author[inst1]{Radoslaw Ryblewski}
\ead{radoslaw.ryblewski@ifj.edu.pl}

\begin{abstract}
A generalized Frenkel condition is proposed for use in spin hydrodynamics to relate the spin density and spin polarization (or spin chemical potential) tensors. It allows for independent treatment of electric- and magnetic-like components of the spin density tensor, which helps to fulfill the stability conditions recently derived in the literature. The generalized Frenkel condition extrapolates between the original Frenkel condition, where only the magnetic-like part of the spin tensor is present, and the case where the spin density tensor is directly proportional to the spin polarization tensor. We also demonstrate that our approach is supported by the result of a microscopic calculation. 
\textcolor{blue}{}
\end{abstract}



\end{frontmatter}



\section{Introduction}

One of the popular assumptions about the form of the spin tensor $S^{\lambda, \,\mu\nu}$ used in spin hydrodynamics is that $S^{\lambda, \,\mu\nu}$ can be entirely expressed by the flow vector $u^\lambda$ and the spin density tensor $S^{\mu \nu}$ \cite{Becattini:2009wh,Florkowski:2018fap}, namely 
\begin{equation}
S^{\lambda, \,\mu\nu} = u^\lambda S^{\mu \nu}.
\label{eq:phform}
\end{equation}
This ansatz can be traced back to the seminal papers by Weyssenhoff and Rabbe~\cite{Weyssenhoff:1947iua,Weyssenhoff:1947vye}, where the first version of spin hydrodynamics was formulated. Nowadays, the form (\ref{eq:phform}) is often called the {\it phenomenological} expression for the spin tensor, to contrast it with other more complex forms resulting from microscopic classical and quantum calculations \cite{Weickgenannt:2022zxs,Becattini:2023ouz}. 

Recent studies of stability and causality properties of spin hydrodynamics~\cite{Hattori:2019lfp,Fukushima:2020ucl,She:2021lhe,Gallegos:2021bzp,Hongo:2021ona,Li:2020eon,Montenegro:2017rbu,Peng:2021ago,Ambrus:2022yzz,Cao:2022aku,Hu:2022azy,Kiamari:2023fbe,Daher:2022xon,Biswas:2023qsw} have led to a somewhat disturbing conclusion that the electric-like components of the spin density tensor, $S^{0i}$, and its magnetic-like components, $S^{ij}$, should differently depend on the corresponding components of the spin polarization tensor $\omega^{\mu \nu}$\,\footnote{The tensor $\omega^{\mu \nu}$ is sometimes called the spin chemical potential -- we prefer to use the name of the chemical potential for the product $\Omega^{\mu \nu} = T \omega^{\mu \nu}$, where $T$ is the system's temperature. }. To be more specific, following Ref.~\cite{Hattori:2019lfp}, we consider the two susceptibilities
\begin{equation}
\chi_s = \frac{\partial S^{ij}}{\partial \omega^{ij}}, \qquad   \chi_b = \frac{\partial S^{0i}}{\partial \omega^{0i}}. 
\label{eq:chis}
\end{equation}
The stability conditions imply that~\cite{Daher:2022wzf,Sarwar:2022yzs,Xie:2023gbo}:
\begin{equation}
\chi_s > 0, \quad \quad \chi_b < 0 .
\label{eq:stab}
\end{equation}
The problem  with the conditions (\ref{eq:stab}) is that at the same time, one commonly assumes a linear connection between $S^{\mu \nu}$ and $\omega^{\mu \nu}$~\cite{Wang:2021ngp,Biswas:2022bht}
\begin{equation}
S^{\mu \nu}(T,\mu, \omega) = S(T,\mu) \,\omega^{\mu \nu}.
\label{eq:Spropo}
\end{equation}
Here $T$ and $\mu$ denote temperature and (baryon) chemical potential, whereas $S$ is some function of $T$ and $\mu$. Equation~(\ref{eq:Spropo}) implies the same values of $\chi_s$ and $\chi_b$, which contradicts the stability criterion (\ref{eq:stab}). The arguments for using (\ref{eq:Spropo}) are twofold: i) one assumes that $T$, $\mu$, $u^\lambda$, and $\omega^{\mu\nu}$ are fundamental variables of spin hydrodynamics, and ii) for small values of $\omega^{\mu\nu}$ (commonly assumed in the stability analyses), the expansion of $S^{\mu \nu}$ should start with a linear term in $\omega^{\mu \nu}$. 

In this work, we argue that the paradox outlined above can be removed by the observation that the spin polarization tensor can be expressed in terms of two space-like four-vectors, $k$ and $\omega$, and the flow vector $u$ \cite{Florkowski:2017ruc}
\begin{equation}
\omega^{\gamma\delta} = k^\gamma u^\delta 
- k^\delta u^\gamma + \epsilon^{\gamma\delta \rho \sigma} u_\rho \omega_\sigma.
\end{equation}
We note that $k$ and $\omega$ are space-like vectors satisfying the orthogonality conditions $k \cdot u = 0$ and $\omega \cdot u = 0$ (by the way, strictly speaking, $k$ is a vector while $\omega$ is an axial vector). Consequently, the linearity between the components of $S^{\mu \nu}$ and $\omega^{\mu\nu}$ given by Eq.~(\ref{eq:Spropo}) can be naturally generalized to the form
\begin{equation}
S^{\gamma \delta} = S_1 
\left( k^\gamma u^\delta 
- k^\delta u^\gamma \right) + S_2
\epsilon^{\gamma\delta \rho \sigma} u_\rho \omega_\sigma.
\label{eq:Spropogen}
\end{equation}
This expression has the correct Lorentz structure and fulfills the condition that $S^{\gamma \delta} \to 0$ for $\omega^{\gamma \delta} \to 0$. The quantities $S_1$ and $S_2$ are two different functions of $T$ and $\mu$. They can be naturally identified with the functions $\chi_b$ and $\chi_s$ defined in Eqs.~(\ref{eq:chis}). In the local rest frame of the fluid element, where $u^\lambda\!=\!(1,0,0,0)$, we find: $S^{0i} = S_1 \omega^{0i}$ and $S^{ij} = S_2 \omega^{ij}$, hence $S_1 = \chi_b$ and $S_2=\chi_s$. Equivalent expressions to Eq.~(\ref{eq:Spropogen}) are:
\begin{equation}
S^{\gamma \delta} = S_1 
\left( \omega^{\gamma \alpha} u_\alpha u^\delta 
- \omega^{\delta \alpha} u_\alpha u^\gamma \right) -\frac{1}{2} S_2
\epsilon^{\gamma\delta \rho \sigma}  \epsilon_{\tau \beta \alpha \sigma} u_\rho u^\alpha \omega^{\tau \beta},
\label{eq:Spropogeno1}
\end{equation}
\begin{equation}
S^{\gamma \delta} = (S_1 - S_2)
\left( \omega^{\gamma \alpha} u_\alpha u^\delta 
- \omega^{\delta \alpha} u_\alpha u^\gamma \right) + S_2 \omega^{ \gamma\delta}  .
\label{eq:Spropogeno3}
\end{equation}
These formulas reproduce the spin density tensor in terms of the spin polarization tensor and the flow vector. 

In the case $S_1=0 \,\,(\chi_b=0)$, the spin density automatically fulfills the condition $S^{\mu\nu} u_\nu = 0$, known in the literature as the Frenkel (or Weyssenhoff) condition~\cite{Becattini:2009wh,Speranza:2020ilk,DeFalco:2023djo}. Hence, we may think of Eq.~(\ref{eq:Spropogen}) as of a \textit{generalized Frenkel condition} --- a formula that restricts the form of $S^{\gamma \delta}$. It was argued in Ref.~\cite{Daher:2022wzf} that the original Frenkel condition helps to get rid of unstable solutions. In this case, $\chi_b = 0$, which corresponds to neutral stability in the electric sector.

Unfortunately, the use of the original Frenkel condition has also some drawbacks, as it undesirably reduces the number of independent components of the spin density tensor. In this context, one may recall that in global equilibrium the polarization tensor $\omega^{\mu\nu}$ is given by thermal vorticity $\varpi^{\mu\nu}$ with electric and magnetic parts given by the acceleration and vorticity vectors, respectively; see, for instance, Refs.~\cite{Buzzegoli:2017cqy,Shokri:2023rpp,Palermo:2023cup}. In this case, we cannot neglect the electric component. The generalized formula~(\ref{eq:Spropogen}) does not lead to such problems.  Below we show that the conditions $S_1 < 0$ and $S_2 > 0$ are indeed fulfilled in certain kinetic-theory calculations. 

\section{Hydrodynamics with conserved spin -- microscopic example}

One of the popular forms of the spin tensor used in the literature is that introduced by de Groot, van Leeuven, and van Weert in their seminal textbook on the relativistic kinetic theory (denoted in the following with the label GLW)~\cite{DeGroot:1980dk}. In the case where the spin part of the angular momentum is separately conserved, the local equilibrium GLW spin tensor has the structure~\cite{Florkowski:2019qdp}
\begin{equation}
S^{\alpha,\, \beta \gamma}_{\rm GLW} = A_1 u^\alpha \omega^{\beta \gamma} + A_2 u^\alpha u^{[\beta} k^{\gamma]}  
+ A_3 \left( u^{ [\beta} \omega^{\gamma ] \alpha}
+ g^{\alpha [ \beta} k^{\gamma ] }\right),
\label{eq:SGLW}
\end{equation}
where the square brackets denote antisymmetrization $X^{[\alpha \beta]} =   (X^{\alpha \beta} - X^{\beta \alpha})/2$. The coefficients $A_1$, $A_2$, and $A_3$ are functions of temperature and chemical potential. In order to obtain the corresponding spin density tensor $S^{\beta \gamma}_{\rm GLW}$, we consider the projection $u_\alpha S^{\alpha, \,\beta \gamma}_{\rm GLW}$ that gives
\begin{equation}
S^{\beta \gamma}_{\rm GLW} = \left(A_1 - \frac{A_2}{2}  -A_3 \right)
\left( k^\beta u^\gamma - k^\gamma u^\delta \right) + A_1 \epsilon^{\beta \gamma \rho \sigma} u_\rho \omega_\sigma
\label{eq:SGLWden}
\end{equation}
and
%
%
%
\begin{equation}
S^{\alpha,\, \beta \gamma}_{\rm GLW} = u^\alpha S^{\beta \gamma}_{\rm GLW} +   A_3\left(  \Delta^{\alpha[\beta}k^{\gamma]}    +   \epsilon^{\alpha[\beta \lambda \chi} u_\lambda \omega_\chi u^{\gamma]} \right).
\label{eq:SGLW2}
\end{equation}
We note that the second term on the right-hand side of the equation above is explicitly orthogonal to $u$. In Ref.~\cite{Florkowski:2019qdp} the following expressions have been derived for the coefficients $A_1, A_2$, and $A_3$:
\begin{equation}
A_1 = C (n_0 - B_0), \quad A_2 = 2 C (n_0 - 3 B_0),  \quad A_3 = C B_0,  
\end{equation}
where $C = \cosh(\mu/T)$, $B_0 = -2 (T^2/m^2) \sigma_0$, and  $n_0$ ($\sigma_0$) is the equilibrium number density (entropy density) of spinless classical particles with mass $m$ at the temperature $T$. It is easy to notice that $S_1 = A_1- \frac{1}{2}A_2- A_3  = C B_0 < 0$ and $S_2 = A_1 > 0$. Hence,  the properties expected from the stability analysis are fulfilled in the considered microscopic model. In the considered case we find
\begin{equation}
S_1 = - C\, \frac{T^3}{\pi^2} \left[4 K_2(x) + x K_1(x)  \right]
\end{equation}
and

\begin{equation}
S_2 = C\, \frac{T^3}{2 \pi^2} \left[(8+x^2) K_2(x) + 2 x K_1(x)  \right],
\end{equation}
where $K_n(x)$'s are the modified Bessel functions of the second kind and $x=m/T$. For small values of $m/T$, we obtain
\begin{equation}
S_1 = - \frac{8 C T^5}{\pi^2 m^2} , \quad S_2 = \frac{8 C T^5}{\pi^2 m^2}.
\end{equation}
Interestingly, in this limit the two coefficients become exactly opposite, $S_2=-S_1$\,\footnote{We note that we cannot take the limit $m \to 0$ as the formalism of Ref.~\cite{DeGroot:1980dk} has been defined for massive particles.}. On the other hand, in the large mass limit we find
\begin{equation}
S_1 = - \frac{C T^3}{\pi^{3/2}} 
\sqrt{\frac{m}{2T}} e^{-m/T}, 
\quad S_2 =  \frac{C m T^2}{2 \pi^{3/2}} 
\sqrt{\frac{m}{2T}} e^{-m/T}.
\end{equation}
In this case $S_1 \ll S_2$ and the magnetic part dominates the behavior of the spin (density) tensor. This observation agrees with an earlier result, see Eq.~(60) in Ref.~\cite{Florkowski:2017dyn}.

The calculation presented above has been done in a specific formulation of spin hydrodynamics. In the future, it would be interesting to analyze other microscopic approaches~\cite{Hidaka:2022dmn,Gao:2019znl,Hattori:2019ahi,Weickgenannt:2019dks} to verify if similar features also occur there.

\section{Positivity of the contraction \texorpdfstring{$\omega_{\mu \nu} S^{\mu \nu}$}{omega : S}}

In thermodynamic relations used in spin thermodynamics~\cite{Becattini:2009wh,Becattini:2023ouz,Hattori:2019lfp,Fukushima:2020ucl,She:2021lhe,Gallegos:2021bzp,Hongo:2021ona,Li:2020eon,Daher:2022xon,Biswas:2023qsw} one includes the term $\omega_{\mu \nu} S^{\mu \nu}$. With the ansatz (\ref{eq:Spropo}), it gives a contribution of indefinite sign (note that $k^2 <0$ and $\omega^2 <0$)
\begin{equation}
\omega_{\mu \nu} S^{\mu \nu}  
= S \omega_{\mu \nu} \omega^{\mu \nu} 
=  -2 S \left(\omega^2 -  k^2 \right).
\end{equation}
Using the generalized Frenkel condition (\ref{eq:Spropogen}), we obtain 
\begin{equation}
\omega_{\mu \nu} S^{\mu \nu}  =  2 S_1  k^2 -2 S_2 \omega^2 .
\end{equation}
With $S_1 < 0$ and $S_2 > 0$ this is a manifestly positive quantity. We conclude that using Eq.~(\ref{eq:Spropogen}) one may avoid some earlier problems found in the literature~\cite{Florkowski:2017ruc} connected with the calculation of the expression like $\sqrt{\omega_{\mu \nu} \omega^{\mu \nu}}$, provided it can be suitably replaced by $\sqrt{\omega_{\mu \nu} S^{\mu \nu}}$.

\section{Summary}

In this work, we have proposed a new form of the relation connecting the spin density tensor $S^{\mu \nu}$ with the spin polarization tensor $\omega^{\mu \nu}$. It may be considered a generalized Frenkel condition. Alternatively, we may regard the new formula (\ref{eq:Spropogen}) as a special form of the spin equation of state used in spin hydrodynamics that guarantees the stability of perturbations around the uniform background. In this way, we have solved problems that have plagued the stability analyses of spin hydrodynamics in recent years. Suggestions that the stability of spin hydrodynamics may depend on the form of the spin equation of state were discussed earlier in Ref.~\cite{Xie:2023gbo}, however, no explicit solution to this problem was given.

We stress that our discussion refers to the calculations based on the ansatz (\ref{eq:phform}) and implies that Eq.~(\ref{eq:Spropo}) should be replaced by Eq.~(\ref{eq:Spropogen}). As our microscopic example shows, the spin tensor may also have parts transverse to the flow vector; see the last term in Eq.~(\ref{eq:SGLW}). The stability studies of spin hydrodynamics with more complicated forms of the spin tensor should definitely be done in the future to achieve more general conclusions on stability. Finally, the stability conditions discussed here refer to low-momentum modes. It would be interesting to extend this study to large momenta.

\medskip
{{\bf Acknowledgements:}} We thank David Wagner for illuminating discussions. A.D. gratefully acknowledges fruitful discussions with Francesco Becattini, Shi Pu, Dong-Lin Wang, and Arpan Das. This work was supported in part by the Polish National Science Centre Grants Nos. 2018/30/E/ST2/00432 (AD, RR) and 2022/47/B/ST2/01372 (WF).

 \bibliographystyle{elsarticle-num} 
 \bibliography{bibliography}
\end{document}